\documentclass[conference]{IEEEtran}

\usepackage{float}

\ifCLASSINFOpdf
   \usepackage[pdftex]{graphicx}
\else
\fi

\begin{document}

\title{Performance evaluation for CRUD operations in asynchronously replicated document oriented database}

\author{
	\IEEEauthorblockN{
		Ciprian-Octavian Truic\u{a}$^{*, 1}$,
		Florin R\u{a}dulescu$^{*, 2}$, 
		Alexandru Boicea$^{*, 3}$
		and Ion Bucur$^{*, 4}$
	}
	\IEEEauthorblockA{
		$^{*}$Department of Computer Science, Faculty of Automatic Control and Computers,\\University Politehnica of Bucharest, Bucharest, Romania
	}
	\IEEEauthorblockA{
		 $^{1}$ciprian.truica@cs.pub.ro, $^{2}$florin.radulescu@cs.pub.ro, $^{3}$alexandru.boicea@cs.pub.ro, $^{4}$ion.bucur@cs.pub.ro
	}
}


%


\maketitle

\begin{abstract}
NoSQL databases are becoming increasingly popular as more developers seek new ways for storing information. The popularity of these databases has risen due to their flexibility and scalability needed in domains like Big Data and Cloud Computing. This paper examines asynchronous replication, one of the key features for a scalable and flexible system. Three of the most popular Document-Oriented Databases, MongoDB, CouchDB, and Couchbase, are examined. For testing, the execution time for CRUD operations for a single database instance and for a distributed environment with two nodes is taken into account and the results are compared with tests outcomes obtained for three relational database management systems: Microsoft SQL Server, MySQL, and PostgreSQL. 
\end{abstract}



%
\IEEEpeerreviewmaketitle

\section{Introduction}

NoSQL is a generic name for database management systems not aligned to the relational model widely used by the industry. One key aspect that differentiates NoSQL databases from the relational ones is that tables and SQL language are not always used. Another key feature is that these databases are optimized for CREATE and READ operations and, often, they offer reduced functionality for UPDATE and DELETE queries. Although this could be seen as a loss in flexibility, NoSQL DBMSs gain in terms of scalability and are more efficient when specific data models are used \cite{IEEEhowto:r1}.

Also, NoSQL systems are designed to work with large amounts of data that does not require a rigid schema, but a more flexible one. Although information can be structured, the emphasis is not on data structure, but on the way data is stored, accessed and processed. This approach is very useful when working with large volumes of information that are processed intensively and analyzed in real time \cite{IEEEhowto:r2}.

NoSQL databases are often classified either by taking into account the way data is organized or by the way data is stored \cite{IEEEhowto:r1}. Each NoSQL database organizes and groups data in different ways, e.g. collections of documents for document oriented databases, associative arrays for key-value databases, etc. Data, in document oriented databases, is stored as collections of documents \cite{IEEEhowto:r3}. Compared to relational databases, collections can be considered analogue to tables and documents analogue to records \cite{IEEEhowto:r4}. The three databases used to test asynchronous replication can be classified by taking into account the way data is stored, the persistence design \cite{IEEEhowto:r5}, as following: MongoDB uses B-trees \cite{IEEEhowto:r6}, CouchDB uses append-only B-trees \cite{IEEEhowto:r7} and Couchbase uses memory with checkpoints \cite{IEEEhowto:r8}. 

One of the key features of NoSQL systems is that they manage data distributed across multiple sites \cite{IEEEhowto:r9}. NoSQL databases were designed to fit with the highly distributed nature of the three-tier internet architecture, and so, due to the persistence design and data structure, they can be easily partitioned on different machines \cite{IEEEhowto:r7}.

The most common distributed model used for distributed databases is replication. There are three types of replication models: asynchronous replication, synchronous replication and semi-synchronous replication \cite{IEEEhowto:r6}\cite{IEEEhowto:r7}\cite{IEEEhowto:r8}\cite{IEEEhowto:r20}. This paper deals with implementing and testing asynchronous replication for document oriented databases. Tests were designed to compute the execution time for CRUD operation. Each test uses a CRUD command to modify the database, and stops when the database is in the same state on all the nodes of the distributed environment. Replication is automatically triggered, and parameter values for replication status are checked if all the database nodes are synchronized. To compare performance time, tests were done on single instances of the document oriented database. The execution time is also compared with the results from similar tests obtained in \cite{IEEEhowto:r10} for three relational systems. The distributed environment uses a master-slave architecture composed of two virtual machines with the same hardware specifications. Even though the tested databases have the same data model and the same data representation (JSON), there are significant differences regarding execution time for CRUD operations for single instance DBMS and for distributed DBMS.

This paper is structured as follows. Section II discusses related work done for comparing different types of NoSQL databases. Section III presents document oriented databases and some key features for the three NoSQL database management systems tested: MongoDB, CouchDB and Couchbase. Section IV discuses our experimented setup in detail. Section V presents experimental data and findings and discusses the results. The final section concludes with a summary.

\section{Related Work}

Todd Hoff presented a NoSQL taxonomy based on the data model \cite{IEEEhowto:r11} and a similar taxonomy which is less fine-grained and comprehensive is done by Ken North \cite{IEEEhowto:r12}. Other classification is done by Ben Scofield \cite{IEEEhowto:r3}.

Another feature of NoSQL systems used for evaluation is the scalability. Jonathan Ellis investigates in his work only the write operation in a replicated environment, because, he argues, it is easy to scale read operations \cite{IEEEhowto:r5}. In his work he excludes some NoSQL database systems because they are not distributed in the way he requires \cite{IEEEhowto:r1}. 

Persistence design was also taken into account to differentiate and compare different types of such databases. In his work, Christof Strauch classifies NoSQL databases into three categories: In-Memory Databases, Memtables and SSTables and B-trees \cite{IEEEhowto:r1}. Some databases use as persistence model, either one of these designs or a combination of different persistence models (e.g. CouchDB uses append-only B-Tree, Couchbase uses memcache with disk persistence, etc.).

Regarding the data model, some literature presents a comparison between the SQL four languages – DML, DDL, DCL and TCL – and the JavaScript API used in MongoDB to create and manipulate information \cite{IEEEhowto:r13}.

Performance evaluation was done by Alexandru Boicea et al., where a performance time comparison for CRUD operations between MongoDB and Oracle database \cite{IEEEhowto:r14}. Yishan Li and Sathiamoorthy Manoharan \cite{IEEEhowto:r15} present in their work a well researched performance comparison between different types of key-value databases.

NoSQL databases began to be used in Cloud Computing because the systems meet the requirements for flexibility and scalability. The old model of centralizing data stores in one location is starting to change: increasingly more applications use decentralized data stores because they provide good mechanisms for fail over, removing the single point of failure, and due to their scalability and flexibility \cite{IEEEhowto:r16}.

NoSQL systems are also used with Big Data because processing can be conveniently done using a distributed system instead of a centralized system. Analyzing data on a replica will not hinder other systems that use the same database.

The motivation behind this paper is to analyze three main document oriented databases \cite{IEEEhowto:r17} and compare their performance time, when working with large amounts of data, in order to determine which database is best suited for systems that process and analyze documents in a distributed environment. This is a Big Data in Cloud Computing problem.

\section{Document Oriented Databases}

Document oriented databases started as a subclass of key-value databases. In time, they have become a NoSQL database class on its own due to the fact that they can store more complex data types. Also, unlike key-value databases, document oriented databases generally support secondary indexes \cite{IEEEhowto:r13}.

These database management systems were specially developed to store, manage and process data using a semi-structured model. Just as relational databases were built around the concepts of relational algebra, document oriented databases have been developed around the notion of document. Documents are encapsulated in collections of documents \cite{IEEEhowto:r18}. In computer science a document is used to describe a text file that has a structure and a design, a file format – a standard method used for encrypting and storing information. A document is stored in a sustainable virtual environment, being available to different applications. Document oriented databases generally encapsulate and encrypt data using standard formats, the most popular being XML, YAML, JSON, BSON, PDF and Microsoft Word documents. A flexibility feature related to the way data is stored is that a document can contain other nested documents \cite{IEEEhowto:r2}. CouchDB and Couchbase use the JSON format for storing data, and MongoDB uses BSON. BSON, or binary JSON, offers support for data types and for encapsulating arrays and nested documents.

Storing and manipulating information in document oriented databases may be similar, in some respects, to the way relational databases handle data, namely a collection can be considered similar to a table and a document can be considered similar to a record. However, in relational databases, data must comply with a fixed schema that structures the record attributes. Unlike relational databases, document oriented records can have a variable number of attribute fields, and if an attribute value is missing, then the field can be omitted. These features alone reinforce the concept of a schema-free database.

Document oriented databases allow the use of keys: in a collection each document has a unique key, similar with a primary key in a relational database, used to uniquely identify a document and to improve CRUD operations performance. MongoDB uses a unique id, \_id, for identifying a document \cite{IEEEhowto:r6}. In CouchDB and Couchbase there is another unique id, besides the \_id key, used for identifying the latest version of the record and for keeping track of the record revisions \cite{IEEEhowto:r7}\cite{IEEEhowto:r8}.

Other key features of document oriented databases are the ability to horizontally scale CRUD operations throughput and to replicate and distribute data over many servers \cite{IEEEhowto:r19}. The design of the distributed architecture for this class of databases is called share nothing horizontal scaling, enabling support for processing a large number of CRUD operations. Shared nothing architecture refers to the fact that the server's resources are not shared; each server manages its own resources. Couchbase can also use vertical scaling, sharing the entire cluster nodes resources between the servers, but this approach is not discussed in this paper, the main focus being horizontal scaling. 

All the tested document oriented databases use asynchronous replication. MongoDB replication is based on Replica Sets, a group of processes providing redundancy and high availability \cite{IEEEhowto:r6}. Replication in CouchDB is incremental: each modification will have a new version \cite{IEEEhowto:r7}. Couchbase supports two types of replication. The first one is replication for the same cluster; the second one is called XDCR (Cross Data Center Replication) and is used to replicate date between clusters. This second type of replication was used for testing\cite{IEEEhowto:r8}.

\section{Experimental Setup}

The experimental setup consists of nine virtual machines created using VMware 10. All the virtual machines have the same hardware configuration: 2GB RAM, 1 CPU with two 2.8GHz cores, 1 60GB HDD and 1 Network Adapter. They all reside in the same LAN and have a static network configuration. The operating system used to test these databases was Ubuntu 14.04LST x64. For each database three virtual machines where created as follows: one for the single instance and two for the replicated architecture. 

\begin{figure}[H]
\centering
\includegraphics[width=2.5in]{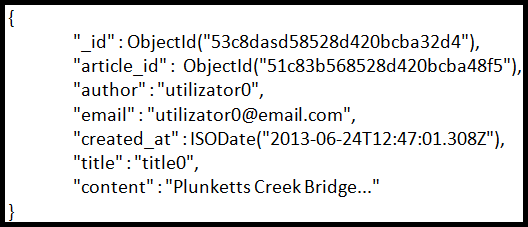}
\caption{Comment collection document sample}
\label{fig1}
\end{figure}

The database structure stores blog posts, and contains two collections: the articles collection and the comments collection. The articles collection stores information about an article and it was not be used for performance tests. The comments collection stores 1KB documents containing comments for a given article identified by its id. Figure~\ref{fig1} presents a document sample. 

The databases comparison involved testing performance time for all the CRUD operations, first on the single instance and then in the distributed environment. The experiments tested the performance of:
\begin{enumerate}
\item	CREATE. New data is added to the database using JavaScript for MongoDB and the Rest API for CouchDB and Couchbase. Records are added one at a time.
\item	READ. This selects and returns all the comments for a given article. This operation is done on the single instance only, because in the distributed Master-Slave configuration only one machine executes the read.
\item	UPDATE. The tests designed for this operation update the content field of a document in the comments collection with new data. For MongoDB this is done by directly updating the field. For CouchDB and Couchbase it required a SELECT operation that gets all the information and then UPDATE can be performed. This is because the UPDATE is at document level \cite{IEEEhowto:r7}\cite{IEEEhowto:r8}. Before update, all the documents are scanned for getting the unique identifier, but this operation is not accounted when calculating the performance time.
\item	DELETE. This operation deletes all the data in the database.
\end{enumerate}

Testing was done using batch processing, in this order: insert, update, select, delete. Each batch of tests was run 50 times for 1000, 10000, and 100000 records. For the distributed environment, insert, update and delete operations are considered finished if the status of the replica is in the same state as the master. This is done by checking replication status parameters available at the DBMS level. Replication is triggered automatically when data is added or modified on the master node. Both nodes are always up and running. If one of the nodes became unavailable, tests are resumed when the distributed system works properly - both nodes are available.

Table~\ref{table1} illustrates the software versions installed for testing. The relational systems use a schema similar to the document oriented databases and test the same CRUD operations \cite{IEEEhowto:r10}.

\begin{table}[H]
\caption{Database Versions}
\label{table1}
\centering
\begin{tabular}{|c|c|}
\hline
\textbf{Database}         & \textbf{Version} \\ \hline
MongoDB                   & 2.4.4            \\ \hline
CouchDB                   & 1.5.0            \\ \hline
Couchbase                 & 2.2.0            \\ \hline
Microsoft SQL Server 2012 & 11.0.3153        \\ \hline
MySQL                     & 5.5.31           \\ \hline
PostgreSQL                & 9.1              \\ \hline
\end{tabular}
\end{table}

\section{Experimental Results}

\begin{figure}[H]
\centering
\includegraphics[width=3in]{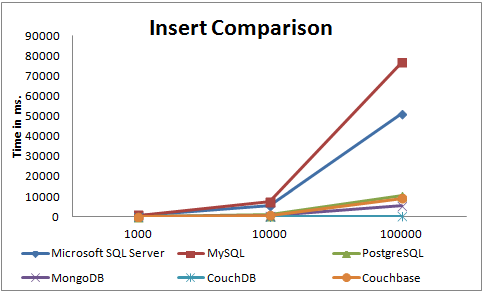}
\caption{Mean performance time in milliseconds for the insert operation}
\label{fig2}
\end{figure}

Figure~\ref{fig2} presents the average execution time of the INSERT operation on the single instance. It shows that the NoSQL systems have the best performances for this operation, with the fastest result given by CouchDB. The performance time of Couchbase is very similar with the performance time of PostgreSQL. There is a difference of a factor of almost 10 between the performance time of Microsoft SQL Server and MongoDB.

\begin{table}[H]
\caption{Mean performance time in milliseconds for the insert operation}
\label{table2}
\centering
\begin{tabular}{c|c|c|c|}
\cline{2-4}
\textbf{}                                  & \multicolumn{3}{c|}{\textbf{Number of operations}}                          \\ \hline
\multicolumn{1}{|c|}{\textbf{Database}}    & \textit{\textbf{1000}} & \textit{\textbf{10000}} & \textit{\textbf{100000}} \\ \hline
\multicolumn{1}{|c|}{Microsoft SQL Server} & 530.1                  & 5516.20                 & 51075.7                  \\ \hline
\multicolumn{1}{|c|}{MySQL}                & 757.1                  & 7326.4                  & 76705.7                    \\ \hline
\multicolumn{1}{|c|}{PostgreSQL}           & 80.9                   & 798.7                   & 10476.7                   \\ \hline
\multicolumn{1}{|c|}{MongoDB}              & 54.9                   & 533.8                   & 5282.5                     \\ \hline
\multicolumn{1}{|c|}{CouchDB}              & 1,39                   & 19,7                    & 141,95                   \\ \hline
\multicolumn{1}{|c|}{Couchbase}            & 77,5                   & 783,67                  & 9188,13                   \\ \hline
\end{tabular}
\end{table}

\begin{figure}[H]
\centering
\includegraphics[width=3in]{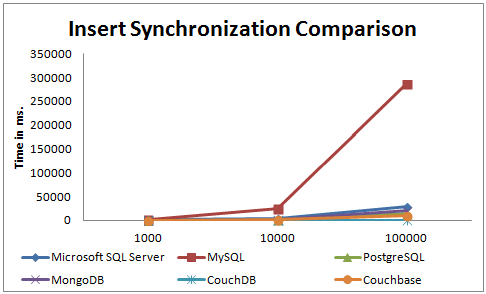}
\caption{Mean performance time in milliseconds for the insert operation for the replicated environment}
\label{fig3}
\end{figure}

In the distributed environment (Figure~\ref{fig3}), CouchDB has the lowest average time for the INSERT operation. MongoDB performs slightly worse than PostgreSQL for a high number of operations. Again, the average performance time for Couchbase and PostgreSQL are very close.

\begin{table}[H]
\caption{Mean performance time in milliseconds for the insert operation for the replicated environment}
\label{table3}
\centering
\begin{tabular}{c|c|c|c|}
\cline{2-4}
\textbf{}                                  & \multicolumn{3}{c|}{\textbf{Number of operations}}                          \\ \hline
\multicolumn{1}{|c|}{\textbf{Database}}    & \textit{\textbf{1000}} & \textit{\textbf{10000}} & \textit{\textbf{100000}} \\ \hline

\multicolumn{1}{|c|}{Microsoft SQL Server} 	& 1000   & 3000    & 28000    \\ \hline
\multicolumn{1}{|c|}{MySQL}					& 2438.5 & 25465.1 & 286934.4 \\ \hline
\multicolumn{1}{|c|}{PostgreSQL}           & 111.2  & 1191.3  & 13429.1  \\ \hline
\multicolumn{1}{|c|}{MongoDB}              & 86.7   & 914     & 19532.3  \\ \hline
\multicolumn{1}{|c|}{CouchDB}              & 2.47   & 22.25   & 215.55   \\ \hline
\multicolumn{1}{|c|}{Couchbase}            & 77.59  & 783.67  & 9188.13  \\ \hline
\end{tabular}
\end{table}

\begin{figure}[H]
\centering
\includegraphics[width=3in]{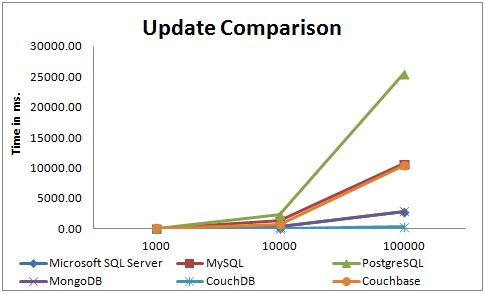}
\caption{Mean performance in milliseconds time for the update operation}
\label{fig4}
\end{figure}

\begin{figure}[H]
\centering
\includegraphics[width=3in]{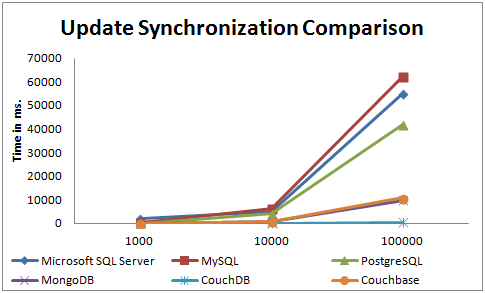}
\caption{Mean performance time in milliseconds for the update operation for the replicated environment}
\label{fig5}
\end{figure}

The UPDATE average time (Figure~\ref{fig4}) for a single instance is lower than the average time for insert regarding Microsoft SQL Server; they are very close to the performance time calculated for MongoDB. PostgreSQL has the worst performance. The update time for MySQL is very close to the update time of Couchbase.  For the distributed architecture (Figure~\ref{fig5}) the NoSQL systems have the best performance time for this operation. 

\begin{table}[H]
\caption{Mean performance in milliseconds time for the update operation}
\label{table4}
\centering
\begin{tabular}{c|c|c|c|}
\cline{2-4}                                           & \multicolumn{3}{c|}{\textbf{Number of operations}}                          \\ \hline
\multicolumn{1}{|c|}{\textbf{Database}}    & \textit{\textbf{1000}} & \textit{\textbf{10000}} & \textit{\textbf{100000}} \\ \hline
\multicolumn{1}{|c|}{Microsoft SQL Server} & 36.10                  & 286.50                  & 2764.80                  \\ \hline
\multicolumn{1}{|c|}{MySQL}                & 87.7                   & 1264                    & 10620.5                  \\ \hline
\multicolumn{1}{|c|}{PostgreSQL}           & 77.3                   & 2385.2                  & 25421.5                  \\ \hline
\multicolumn{1}{|c|}{MongoDB}              & 17.3                   & 265.4                   & 2875.9                   \\ \hline
\multicolumn{1}{|c|}{CouchDB}              & 1.56                   & 18.64                   & 266.68                   \\ \hline
\multicolumn{1}{|c|}{Couchbase}            & 73.16                  & 731.39                  & 10414.85                 \\ \hline
\end{tabular}
\end{table}

\begin{table}[H]
\caption{Mean performance time in milliseconds for the update operation for the replicated environment}
\label{table5}
\centering
\begin{tabular}{c|c|c|c|}
\cline{2-4}
                                           & \multicolumn{3}{c|}{\textbf{Number of operations}}                          \\ \hline
\multicolumn{1}{|c|}{\textbf{Database}}    & \textit{\textbf{1000}} & \textit{\textbf{10000}} & \textit{\textbf{100000}} \\ \hline
\multicolumn{1}{|c|}{Microsoft SQL Server} & 2000                   & 5000                    & 55000                    \\ \hline
\multicolumn{1}{|c|}{MySQL}                & 467.3                  & 6265.4                  & 62396                    \\ \hline
\multicolumn{1}{|c|}{PostgreSQL}           & 88.9                   & 3932.5                  & 42020.3                  \\ \hline
\multicolumn{1}{|c|}{MongoDB}              & 54.4                   & 583.2                   & 9741.7                   \\ \hline
\multicolumn{1}{|c|}{CouchDB}              & 2.64                   & 22.96                   & 391.55                   \\ \hline
\multicolumn{1}{|c|}{Couchbase}            & 79.6                   & 779.01                  & 10737.05                 \\ \hline
\end{tabular}
\end{table}

\begin{figure}[H]
\centering
\includegraphics[width=3in]{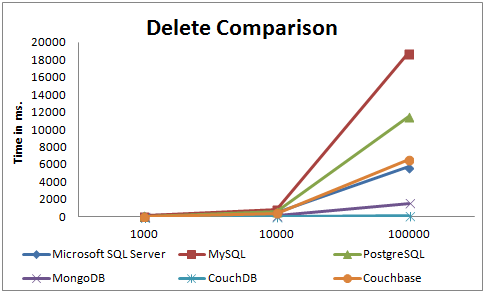}
\caption{Mean performance time in milliseconds for the delete operation}
\label{fig6}
\end{figure}

\begin{figure}[H]
\centering
\includegraphics[width=3in]{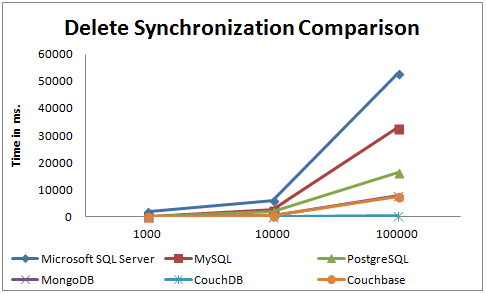}
\caption{Mean performance time in milliseconds for the delete operation for the replicated environment}
\label{fig7}
\end{figure}

Among the relational systems, Microsoft SQL Server has the best performance time for the DELETE operation in single instance architecture (Figure~\ref{fig6}) and the worst performance time in the replicated environment (Figure~\ref{fig7}). For the document oriented databases, again CouchDB has the lowest time in both cases, followed by MongoDB and Couchbase.

\begin{figure}[H]
\centering
\includegraphics[width=3in]{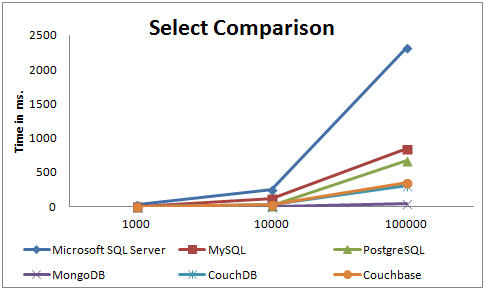}
\caption{Mean performance time for the select operation}
\label{fig8}
\end{figure}

The SELECT operation is faster for the document oriented systems than for the relational systems when performing a large number of operations (Figure~\ref{fig8}).

\begin{table}[H]
\caption{Mean performance time in milliseconds for the delete operation}
\label{table6}
\centering
\begin{tabular}{c|c|c|c|}
\cline{2-4}
                                           & \multicolumn{3}{c|}{\textbf{Number of operations}}                          \\ \hline
\multicolumn{1}{|c|}{\textbf{Database}}    & \textit{\textbf{1000}} & \textit{\textbf{10000}} & \textit{\textbf{100000}} \\ \hline
\multicolumn{1}{|c|}{Microsoft SQL Server} & 127                    & 482.90                  & 5715.4                   \\ \hline
\multicolumn{1}{|c|}{MySQL}                & 78.3                   & 825.8                   & 18794.4                  \\ \hline
\multicolumn{1}{|c|}{PostgreSQL}           & 35.5                   & 582.6                   & 11479.8                  \\ \hline
\multicolumn{1}{|c|}{MongoDB}              & 9                      & 133.8                   & 1530.9                   \\ \hline
\multicolumn{1}{|c|}{CouchDB}              & 1.19                   & 15.57                   & 132.7                    \\ \hline
\multicolumn{1}{|c|}{Couchbase}            & 39.37                  & 405.57                  & 6579.23                  \\ \hline
\end{tabular}
\end{table}

\begin{table}[H]
\caption{Mean performance time in milliseconds for the delete operation for the replicated environment}
\label{table7}
\centering
\begin{tabular}{c|c|c|c|}
\cline{2-4}
                                           & \multicolumn{3}{c|}{\textbf{Number of operations}}                          \\ \hline
\multicolumn{1}{|c|}{\textbf{Database}}    & \textit{\textbf{1000}} & \textit{\textbf{10000}} & \textit{\textbf{100000}} \\ \hline
\multicolumn{1}{|c|}{Microsoft SQL Server} & 2000                   & 6000                    & 53000                    \\ \hline
\multicolumn{1}{|c|}{MySQL}                & 112                    & 2529.8                  & 32897.9                  \\ \hline
\multicolumn{1}{|c|}{PostgreSQL}           & 86.8                   & 2040.6                  & 16465.2                  \\ \hline
\multicolumn{1}{|c|}{MongoDB}              & 26.4                   & 321.4                   & 7814.6                   \\ \hline
\multicolumn{1}{|c|}{CouchDB}              & 2.1                    & 18.73                   & 227.22                   \\ \hline
\multicolumn{1}{|c|}{Couchbase}            & 45.83                  & 445.72                  & 7578.98                  \\ \hline
\end{tabular}
\end{table}

\begin{table}[H]
\caption{Mean performance time for the select operation}
\label{table8}
\centering
\begin{tabular}{c|c|c|c|}
\cline{2-4}

                                           & \multicolumn{3}{c|}{\textbf{Number of operations}}                          \\ \hline
\multicolumn{1}{|c|}{\textbf{Database}}    & \textit{\textbf{1000}} & \textit{\textbf{10000}} & \textit{\textbf{100000}} \\ \hline
\multicolumn{1}{|c|}{Microsoft SQL Server} & 35.3                   & 243.6                   & 2313.4                   \\ \hline
\multicolumn{1}{|c|}{MySQL}                & 4.1                    & 117.8                   & 844.8                    \\ \hline
\multicolumn{1}{|c|}{PostgreSQL}           & 3.7                    & 19.4                    & 663.5                    \\ \hline
\multicolumn{1}{|c|}{MongoDB}              & 1                      & 6                       & 43.5                     \\ \hline
\multicolumn{1}{|c|}{CouchDB}              & 2.14                   & 30.44                   & 307.54                   \\ \hline
\multicolumn{1}{|c|}{Couchbase}            & 4.34                   & 34.89                   & 345.77                   \\ \hline
\end{tabular}
\end{table}

The results, slightly worse, from Yishan Li and Sathiamoorthy Manoharan \cite{IEEEhowto:r15} for the single instance are close to the results presented in this paper for MongoDB, CouchDB, Couchbase, and Microsoft SQL Server.  In their work they used older versions of the NoSQL databases and the Express Edition for Microsoft SQL Server and they do not state the hardware architecture of the machines. Also they used Java to test the CRUD operations.

The time difference between the relational systems and document oriented systems can arise due to the fact that for the relational case integrity constrains are checked when inserting, updating and deleting records while the NoSQL systems do not check any constraints.

\section{Conclusion}

This paper main focus is to compare the time performance for CRUD operations for different implementations of NoSQL document oriented systems in a distributed environment. Although CouchDB performs very well for the insert, update and delete, MongoDB is the fastest when it comes to fetching data. Overall, the NoSQL databases perform better than the relational ones.

A key feature of MongoDB is the update operation because it is atomic \cite{IEEEhowto:r6}; it is possible to update a particular field as it is done in relational databases. CouchDB and Couchbase have a document update approach, which makes the update operation less flexible. If a NoSQL database that behaves similar to a relational database, but offers better performance is required for developing an application, then MongoDB is a good choice.

Between the document oriented databases CouchDB has an overall very good performance time for the INSERT, UPDATE and DELETE operations, but it falls behind when it comes to modeling the data. This database does not support nested documents \cite{IEEEhowto:r7} as in MongoDB \cite{IEEEhowto:r6} and Couchbase \cite{IEEEhowto:r8}. For applications doing intensive write operations a good choice would be CouchDB.

In both tested environments, the performance of Couchbase and PostgreSQL are very similar. Couchbase main distributed environment is based on a shared everything architecture and every time a new node is added its resources are added to the shared resource pool. This feature makes Couchbase a good choice when it comes to intensive data processing applications.

Database management systems are applications that could change, in time, the way they handle the CRUD operations and this could enhance or degrade their performance. When developing applications it is recommended to make the choice that best fits the requirements after a careful analysis of all features it offers.





%

\end{document}